\documentclass[twocolumn,twocolappendix]{aastex631}
\usepackage{graphicx}
\usepackage{gensymb}
\usepackage{textcomp}
\usepackage{natbib}

\newcommand{\hri}{HRI$_{\textrm{\scriptsize{EUV}}}$}

\begin{document} 

   \title{Fleeting Small-scale Surface Magnetic Fields Build the Quiet-Sun Corona}

\correspondingauthor{L. P. Chitta}
\email{chitta@mps.mpg.de}

\author[0000-0002-9270-6785]{L. P. Chitta}
\affiliation{Max-Planck-Institut f\"ur Sonnensystemforschung, Justus-von-Liebig-Weg 3, 37077 G\"ottingen, Germany}
\author[0000-0002-3418-8449]{S. K. Solanki}
\affiliation{Max-Planck-Institut f\"ur Sonnensystemforschung, Justus-von-Liebig-Weg 3, 37077 G\"ottingen, Germany}
\author[0000-0002-3387-026X]{J. C. del Toro Iniesta}
\affiliation{Instituto de Astrofísica de Andalucía (IAA-CSIC), Apartado de Correos 3004, E-18080 Granada, Spain}
\affiliation{Spanish Space Solar Physics Consortium (S$^3$PC), Spain}
\author{J. Woch}
\affiliation{Max-Planck-Institut f\"ur Sonnensystemforschung, Justus-von-Liebig-Weg 3, 37077 G\"ottingen, Germany}
\author[0000-0003-2755-5295]{D. Calchetti}
\affiliation{Max-Planck-Institut f\"ur Sonnensystemforschung, Justus-von-Liebig-Weg 3, 37077 G\"ottingen, Germany}
\author[0000-0002-9972-9840]{A. Gandorfer}
\affiliation{Max-Planck-Institut f\"ur Sonnensystemforschung, Justus-von-Liebig-Weg 3, 37077 G\"ottingen, Germany}
\author{J. Hirzberger}
\affiliation{Max-Planck-Institut f\"ur Sonnensystemforschung, Justus-von-Liebig-Weg 3, 37077 G\"ottingen, Germany}
\author[0000-0002-4796-9527]{F. Kahil}
\affiliation{Max-Planck-Institut f\"ur Sonnensystemforschung, Justus-von-Liebig-Weg 3, 37077 G\"ottingen, Germany}
\author[0000-0001-7809-0067]{G. Valori}
\affiliation{Max-Planck-Institut f\"ur Sonnensystemforschung, Justus-von-Liebig-Weg 3, 37077 G\"ottingen, Germany}
\author[0000-0001-8829-1938]{D. Orozco Su\'arez}
\affiliation{Instituto de Astrofísica de Andalucía (IAA-CSIC), Apartado de Correos 3004, E-18080 Granada, Spain}
\affiliation{Spanish Space Solar Physics Consortium (S$^3$PC), Spain}
\author[0000-0003-1483-4535]{H. Strecker}
\affiliation{Instituto de Astrofísica de Andalucía (IAA-CSIC), Apartado de Correos 3004, E-18080 Granada, Spain}
\affiliation{Spanish Space Solar Physics Consortium (S$^3$PC), Spain}
\author[0000-0002-1790-1951]{T. Appourchaux}
\affiliation{Univ. Paris-Sud, Institut d’Astrophysique Spatiale, UMR 8617, CNRS, B\^ atiment 121, 91405 Orsay Cedex, France}
\author{R. Volkmer}
\affiliation{Leibniz-Institut für Sonnenphysik, Sch\"oneckstr. 6, D-79104 Freiburg, Germany}
\author[0000-0001-9921-0937]{H. Peter}
\affiliation{Max-Planck-Institut f\"ur Sonnensystemforschung, Justus-von-Liebig-Weg 3, 37077 G\"ottingen, Germany}
\author[0000-0002-7762-5629]{S. Mandal}
\affiliation{Max-Planck-Institut f\"ur Sonnensystemforschung, Justus-von-Liebig-Weg 3, 37077 G\"ottingen, Germany}
\author[0000-0003-1294-1257]{R. Aznar Cuadrado}
\affiliation{Max-Planck-Institut f\"ur Sonnensystemforschung, Justus-von-Liebig-Weg 3, 37077 G\"ottingen, Germany}
\author[0000-0001-7298-2320]{L. Teriaca}
\affiliation{Max-Planck-Institut f\"ur Sonnensystemforschung, Justus-von-Liebig-Weg 3, 37077 G\"ottingen, Germany}
\author[0000-0001-6060-9078]{U. Sch\"{u}hle}
\affiliation{Max-Planck-Institut f\"ur Sonnensystemforschung, Justus-von-Liebig-Weg 3, 37077 G\"ottingen, Germany}
\author[0000-0003-4052-9462]{D. Berghmans}
\affiliation{Solar-Terrestrial Centre of Excellence -- SIDC, Royal Observatory of Belgium, Ringlaan -3- Av. Circulaire, 1180 Brussels, Belgium}
\author[0000-0002-5022-4534]{C. Verbeeck}
\affiliation{Solar-Terrestrial Centre of Excellence -- SIDC, Royal Observatory of Belgium, Ringlaan -3- Av. Circulaire, 1180 Brussels, Belgium}
\author[0000-0002-2542-9810]{A. N. Zhukov}
\affiliation{Solar-Terrestrial Centre of Excellence -- SIDC, Royal Observatory of Belgium, Ringlaan -3- Av. Circulaire, 1180 Brussels, Belgium}
\affiliation{Skobeltsyn Institute of Nuclear Physics, Moscow State University, 119991 Moscow, Russia}
\author[0000-0003-3621-6690]{E. R. Priest}
\affiliation{School of Mathematics and Statistics, University of St Andrews, St Andrews, KY16 9SS, UK}

\received{2023 July 19}
\accepted{2023 August 17}

\submitjournal{ApJL}

\begin{abstract}
Arch-like loop structures filled with million Kelvin hot plasma form the building blocks of the quiet-Sun corona. Both high-resolution observations and magnetoconvection simulations show the ubiquitous presence of magnetic fields on the solar surface on small spatial scales of $\sim$100\,km. However, the question of how exactly these quiet-Sun coronal loops originate from the photosphere and how the magnetic energy from the surface is channeled to heat the overlying atmosphere is a long-standing puzzle. Here we report high-resolution photospheric magnetic field and coronal data acquired during the second science perihelion of Solar Orbiter that reveal a highly dynamic magnetic landscape underlying the observed quiet-Sun corona. We found that coronal loops often connect to surface regions that harbor fleeting weaker, mixed-polarity magnetic field patches structured on small spatial scales, and that coronal disturbances could emerge from these areas. We suggest that weaker magnetic fields with fluxes as low as $10^{15}$\,Mx and/or those that evolve on timescales less than 5\,minutes, are crucial to understand the coronal structuring and dynamics.
\end{abstract}

%
%
\section{Introduction\label{sec:intr}}
The outer atmosphere of the Sun, the corona, is best observed from space in the extreme ultraviolet (EUV) and X-rays. The corona is characterized by the underlying magnetic activity \citep[][]{1974ApJ...189L..93G,1978ARA&A..16..393V}. The so-called active Sun consists of large-scale transient phenomena such as sunspots, active regions, prominences, flux ropes, flares and coronal mass ejections \citep{priest2014}. By comparison, outside the active Sun we have the quiet Sun (QS), which varies generally on smaller scales, and whose corona includes arch-like plasma loops that trace the closed magnetic field \citep[][]{2014LRSP...11....4R} and coronal holes, which appear darker in the EUV and X-rays and are governed by open magnetic fields of the Sun that extend into interplanetary space  \citep[][]{2009LRSP....6....3C}.

This magnetically-closed quiet-Sun corona (simply referred to as the QS corona in the rest of the paper) is also the most persistent outer atmospheric feature and is present throughout the 11-year solar activity cycle, with its global mean temperatures varying from 1.4\,MK to 1.8\,MK with increasing magnetic activity \citep[][]{2017SciA....3E2056M}. Plasma loops in the QS corona have a wide range of lengths from the order of 10\,Mm to more than 100\,Mm \citep[e.g.,][]{2004SoPh..225..227W,2023A&A...673A..81M}. Some QS coronal loops might lack a clear spatial structuring as they appear rather diffuse in the EUV emission \citep[][]{2023arXiv230801640G}. Although it is accepted that magnetic fields govern coronal dynamics in general, the mechanism of coronal heating to over a million Kelvin remains widely debated \citep[][]{2019ARA&A..57..157C}.

Surface magnetic fields, underlying the QS corona, generally reside on spatial scales of less than 100\,km, in the intergranular lanes of turbulent magnetoconvection \citep[][]{1993SSRv...63....1S}. These QS magnetic fields are actually the most dominant form of the solar surface magnetism, with three orders of magnitude more magnetic flux emerging in the quiet regions compared to that in active regions \citep[][]{2012LRSP....9....4S}. Photospheric spectropolarimetric diagnostics based on the Hanle effect indicate the existence of ubiquitous tangled magnetic fields with an average strength over 100\,G in the QS \citep[][]{2004Natur.430..326T,2011ApJ...731L..21S,2019LRSP...16....1B}. A local dynamo process could produce these QS magnetic fields \citep[][]{2007A&A...465L..43V,2010A&A...513A...1D,2011ApJ...737...52L,2014ApJ...789..132R}.

Three-dimensional magnetohydrodynamic models of the solar atmosphere that self-consistently produce QS magnetic fields with average strengths over 100\,G, do lead to an $\sim$1\,MK hot corona \citep[][]{2015Natur.522..188A,2017ApJ...834...10R}. In these models coronal heating is facilitated via the dissipation of Alfv\'en waves with lifetimes of 30--50\,minutes, larger compared to the typical granular lifetimes of 5--8\,minutes \citep[][]{2015Natur.522..188A} -or- through the stressing of magnetic fields with motions on timescales of 20--50\,minutes, fueling intermittent energy release in the corona \citep[][]{2017ApJ...834...10R}. Such models also reproduce features of some observed transient QS coronal brightenings \citep[][]{2021A&A...656L...7C}.

However, the correspondence between these simulations and observations in terms of the spatial structuring and temporal evolution of the QS corona is not fully established. Moreover, observational questions on how these small-scale magnetic fields do couple to the solar corona, along with the understanding of spatial, temporal scales and flux content of magnetic fields relevant for QS coronal structuring and heating remain open. It is often difficult and non-trivial to associate QS atmospheric loops to their surface magnetic footpoints \citep[e.g.,][]{2007A&A...475.1101S}. This is mainly because, until recently there is a lack of joint photospheric and coronal observations at high-spatial resolutions of better than 500\,km (i.e., sub-granular scales where most of the magnetic fields reside in the photosphere). Here we use unprecedented joint photospheric and coronal observations from Solar Orbiter \citep[][]{2020A&A...642A...1M}, during its second science perihelion passage, to study the magnetic landscape of the QS corona.

\begin{figure*}
\begin{center}
\includegraphics[width=\textwidth]{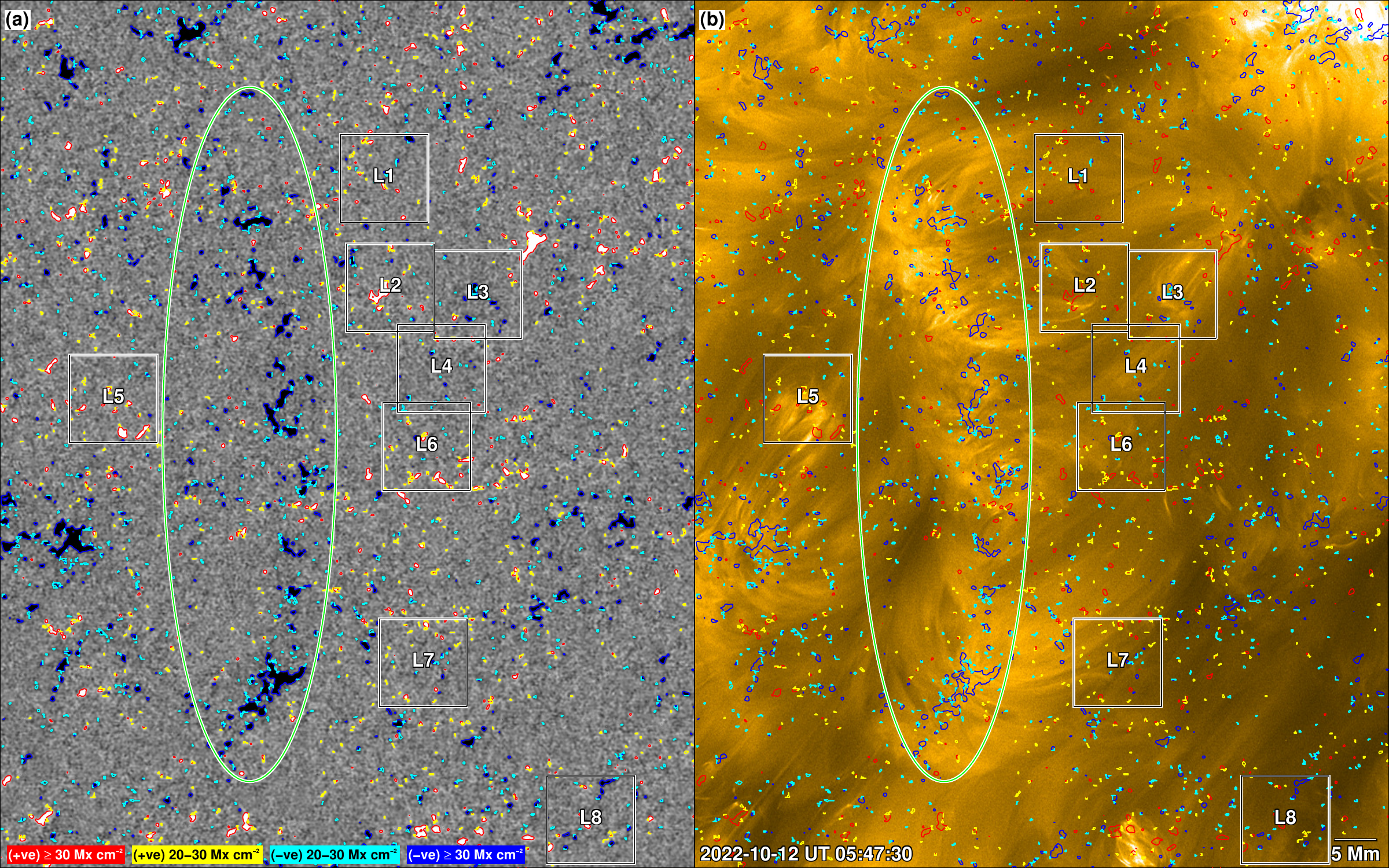}
\caption{Quiet-Sun photosphere and corona. Panel (a): Photospheric magnetic flux density, close to the disk-center, obtained by SO/PHI on 2022 October 12. The lighter and darker shaded regions represent positive and negative magnetic polarity regions, respectively. The map is saturated at flux densities of $\pm$50\,Mx\,cm$^{-2}$. The red (positive) and blue (negative) contours outline magnetic concentrations with flux densities above 30\,Mx\,cm$^{-2}$ (i.e., 3 times the noise level of $\sim$10\,Mx\,cm$^{-2}$). The yellow (positive) and cyan (negative) contours mark regions with flux densities between 20--30\,Mx\,cm$^{-2}$ (i.e., 2--3 times the noise level). The green ellipse outlines a prominent elongated magnetic network feature composed of predominantly negative magnetic polarity concentrations. Boxes L1--L8 (side length of 10.55\,Mm) identify magnetic field patches underlying the loop features that we analyzed in this study. Panel (b): QS coronal features captured by the \hri. The displayed snapshot is from the 30\,s effective cadence time series. The green ellipse, contours, and boxes all have the same meaning as in panel (a). A 5\,Mm scale is overlaid for reference. An animation of panel (b), without the box and ellipse annotations (covering time period UT\,04:40--06:55), is available \href{https://drive.google.com/drive/folders/1pXRVsCwviMlgFMBxzks64qBFpim4SMLV?usp=share_link}{online}. The real-time duration of the animation is 9\,s. See Sect.\,\ref{sec:obs} and Appendix\,\ref{app:proc} for details.\label{fig:over1}}
\end{center}
\end{figure*}

\begin{figure*}
\begin{center}
\includegraphics[width=0.7\textwidth]{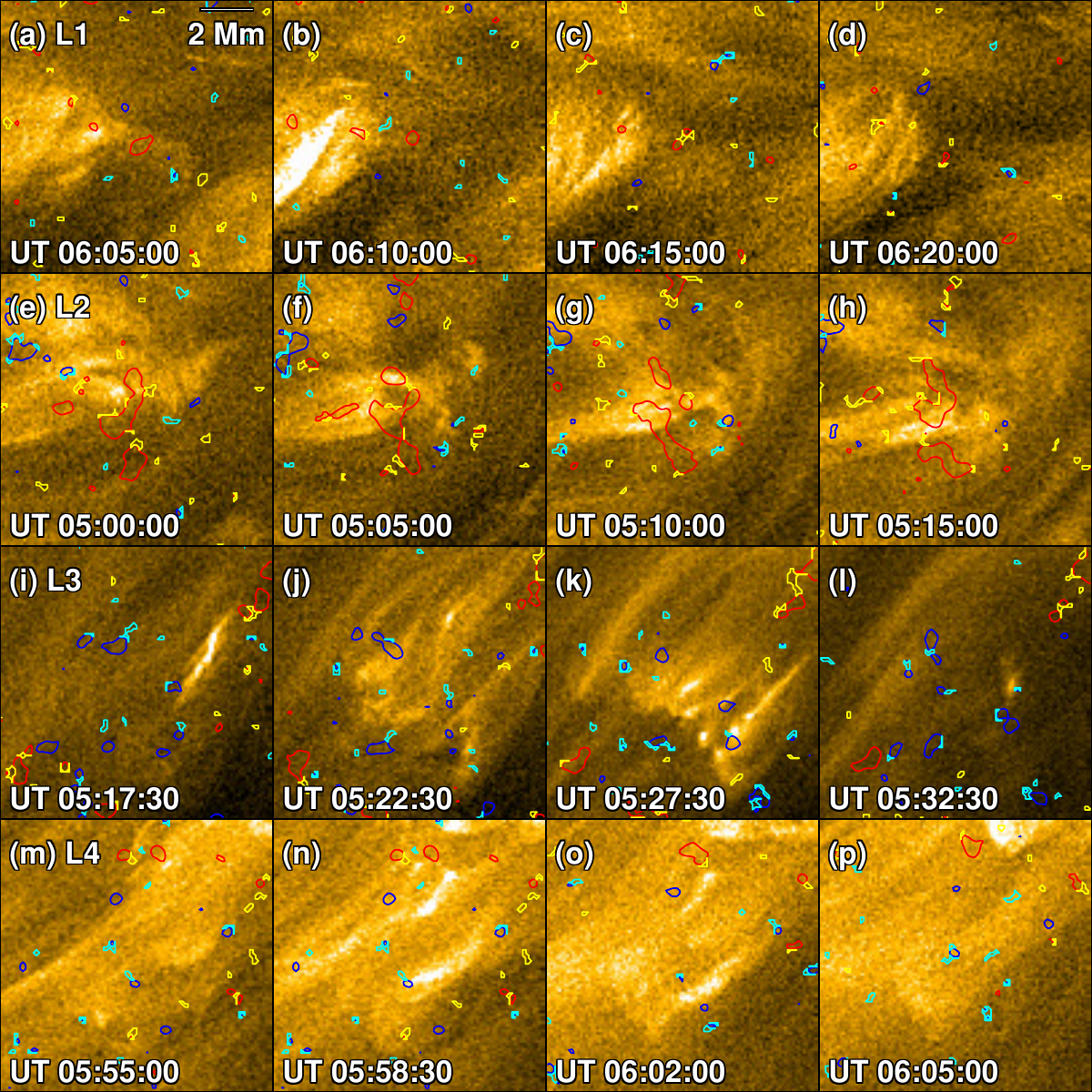}
\caption{Small-scale and weaker magnetic footpoints of QS coronal loops. \hri\ sequences of sections of loops L1--L4, marked in Fig.\,\ref{fig:over1}, and contours of the LOS component of the corresponding magnetic field concentrations at their footpoints, as seen with SO/PHI are displayed. The magnetic contours have the same meaning as in Fig.\,\ref{fig:over1}. Time runs from left to right. See Sect.\,\ref{sec:obs} and Appendix\,\ref{app:proc} for details.\label{fig:lps1}}
\end{center}
\end{figure*}

\section{Observations\label{sec:obs}}
We used photospheric line-of-sight magnetic field or flux density observations and coronal EUV images covering a QS region close to the disk-center. The data were recorded on 2022 October 12, at which time, Solar Orbiter was at a distance of 0.293\,astronomical\,units from the Sun. These observations are a part of the Solar Orbiter Observing Plan, named {\texttt{R\_SMALL\_HRES\_HCAD\_RS-burst}} \citep[][]{2020A&A...642A...3Z}, and are available through the Solar Orbiter Archive.\footnote{https://soar.esac.esa.int/soar}

The magnetic field maps were obtained by the High Resolution Telescope (HRT) of the Polarimetric and Helioseismic Imager on board the Solar Orbiter spacecraft \citep[SO/PHI;][]{2020A&A...642A..11S}, which samples the four Stokes parameters at five wavelength positions within the Fe\,{\sc i} 617.3\,nm line plus a sixth one of its nearby continuum (reduction and calibration of SO/PHI data are detailed in \citealt{2022SPIE12180E..3FK}, \citealt{2022SPIE12189E..1JS}, and \citealt{2023A&A...675A..61K}).\footnote{https://www.mps.mpg.de/solar-physics/solar-orbiter-phi/data-releases} The data were recorded from UT\,04:17 to UT\,07:12, with a cadence of 300\,s, an image scale of 0.5\arcsec\,pixel$^{-1}$ (spatial resolution corresponding to $2\,\textrm{pixels}\approx212\,\textrm{km}$ on the Sun). The HRT field of view (FOV) of 109\,Mm\,$\times$\,109\,Mm is sampled by 1024\,$\times$\,1024\,pixels.

The coronal EUV images were obtained using the High Resolution Imager (\hri) of the Extreme Ultraviolet Imager \citep[EUI;][]{2020A&A...642A...8R} onboard Solar Orbiter. The filter of \hri\ is centered at 17.4\,nm, and its thermal response function has a peak at around 1\,MK \citep[e.g.,][]{2021A&A...656L...7C}. The passband has contributions from Fe\,{\sc ix} (at 17.11\,nm) and Fe\,{\sc x} (at 17.45\,nm\ and 17.72\,nm). In this study we used level-2 data made available as a part of the EUI Data Release 6 \citep[][]{euidatarelease6}. The whole of the \hri\ data sequence may be divided into three sub-sequences. From UT\,04:40 to UT\,05:25, the images were recorded at a cadence of 10\,s with an exposure time of 5.6\,s, then until UT\,06:10 at a cadence of 3\,s with an exposure time of 1.65\,s, and finally again at 10\,s cadence with exposure time of 5.6\,s until about UT\,06:55. The \texttt{EUIPREP} pipeline that produces the level-2 images normalizes the data by the exposure time. Using these three \hri\  sub-sequences, we derived a new series with a uniform effective cadence of 30\,s for display purposes (see Appendix\,\ref{app:proc} for details).

The \hri\ observations have an image scale of 0.492\arcsec\,pixel$^{-1}$ (spatial resolution of $\sim$\,209\,km), and FOV of $\sim$\,214\,Mm\,$\times$\,214\,Mm, sampled by 2048\,$\times$\,2048\,pixels (Fig.\,\ref{fig:euifov}), fully covering the FOV of SO/PHI HRT. We aligned the \hri\ and SO/PHI data, and selected an area of $\sim$\,84\,Mm\,$\times$\,104\,Mm for further analysis. More details on the SO/PHI HRT and \hri\ data processing and alignment are described in the Appendix\,\ref{app:proc}.  

The longitude of Solar Orbiter in the Heliocentric Earth Ecliptic coordinate system at the time of observations was $\sim-118$\textdegree. Therefore, coordinated observations from neither Earth-orbiting nor ground-based telescopes are available.

The photospheric magnetic flux density and the overlying coronal 17.4\,nm emission maps, in the selected FOV are shown in Fig.\,\ref{fig:over1}. As expected, the SO/PHI FOV is dominated by small-scale magnetic fields typical of the QS. The standard deviation ($\sigma$) of the root-mean-square fluctuations of the magnetic flux density is estimated to be about 10\,Mx\,cm$^{-2}$. We consider this to be the 1-$\sigma$ noise level in the inferred magnetic flux densities. Based on this, we classify all those pixels that possess magnetic flux densities above 3-$\sigma$ level, including isolated ones, as regions where the fields are well-detected. Clear magnetic network patches  \citep[known to host kilo Gauss magnetic flux elements;][]{1973SoPh...32...41S}, distributed throughout the FOV are distinguishable (blue and red colored contoured regions in Fig.\,\ref{fig:over1}). One such elongated magnetic network feature composed of predominantly negative polarity magnetic elements is highlighted by an ellipse in Fig.\,\ref{fig:over1}a.

\begin{figure*}
\begin{center}
\includegraphics[width=0.7\textwidth]{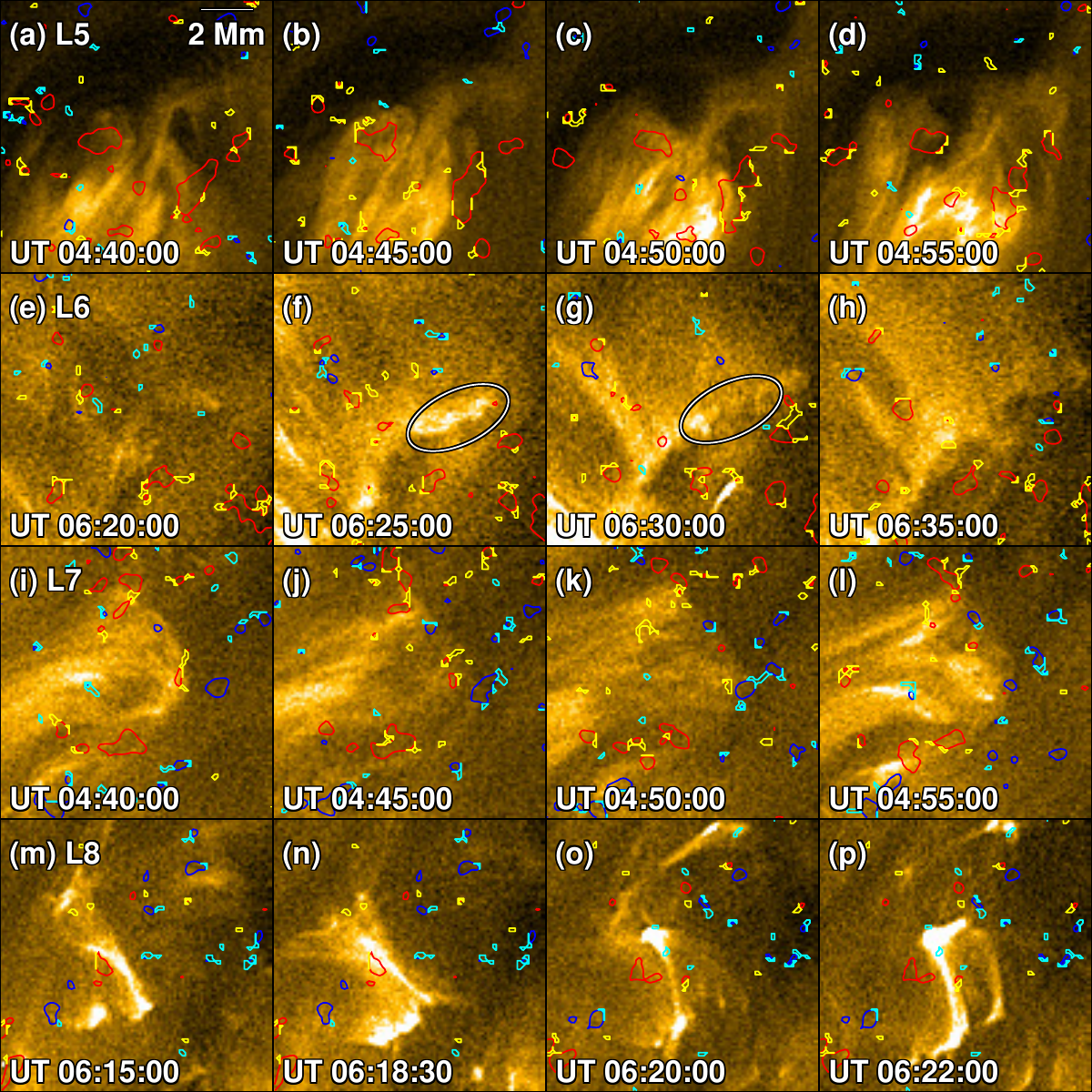}
\caption{Small-scale and weaker magnetic footpoints of QS coronal loops. Same as Fig.\,\ref{fig:lps1}, but plotted for sections of loops L5--L8. Ellipses in panels (f) and (g) outline regions of footpoints of coronal loop bundle L6 without clear underlying magnetic field structures. \label{fig:lps2}}
\end{center}
\end{figure*}

We define weaker field patches as pixels with flux densities in the range of 2-$\sigma$ to 3-$\sigma$ level. In this case, we avoided isolated pixels and considered only those patches which contain more than 4 contiguous pixels, each possessing flux densities in the range of 2-$\sigma$ to 3-$\sigma$ level. We found that these weaker field patches generally surround the network patches (cyan and yellow colored contoured regions in Fig.\,\ref{fig:over1}), nonetheless, there are also clear cases of isolated weaker field patches. In our observations, the well-detected patches represent, 2.9\% of the pixels in the considered FOV on average, while the weaker field patches per our criterion, represent about 1.9\% of the pixels.

The overlying coronal emission is generally structured into arch-like loop features. Many such loops are apparently rooted in the underlying network patches (see loops outlined by the green ellipse in Fig.\,\ref{fig:over1}b). Together, these SO/PHI and \hri\ observations reveal the QS magnetic structures at almost exactly the same high spatial resolution of $\sim$\,210\,km. Using these unprecedented coordinated observations from Solar Orbiter, we now examine the surface magnetic landscape of the QS corona.

\begin{figure*}
\begin{center}
\includegraphics[width=0.49\textwidth]{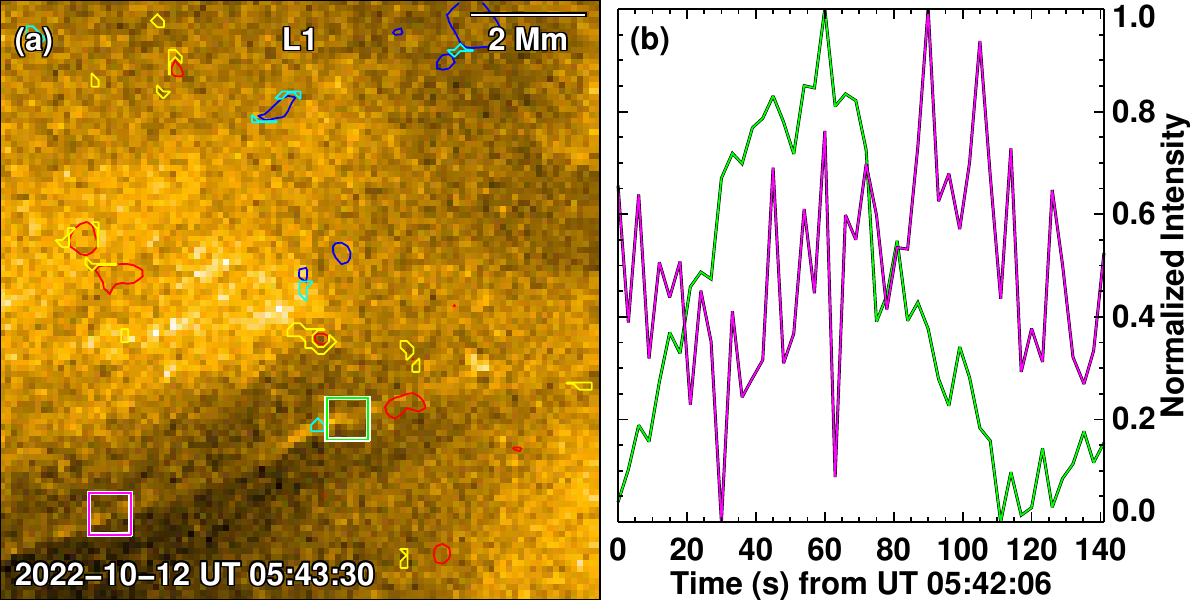}
\includegraphics[width=0.49\textwidth]{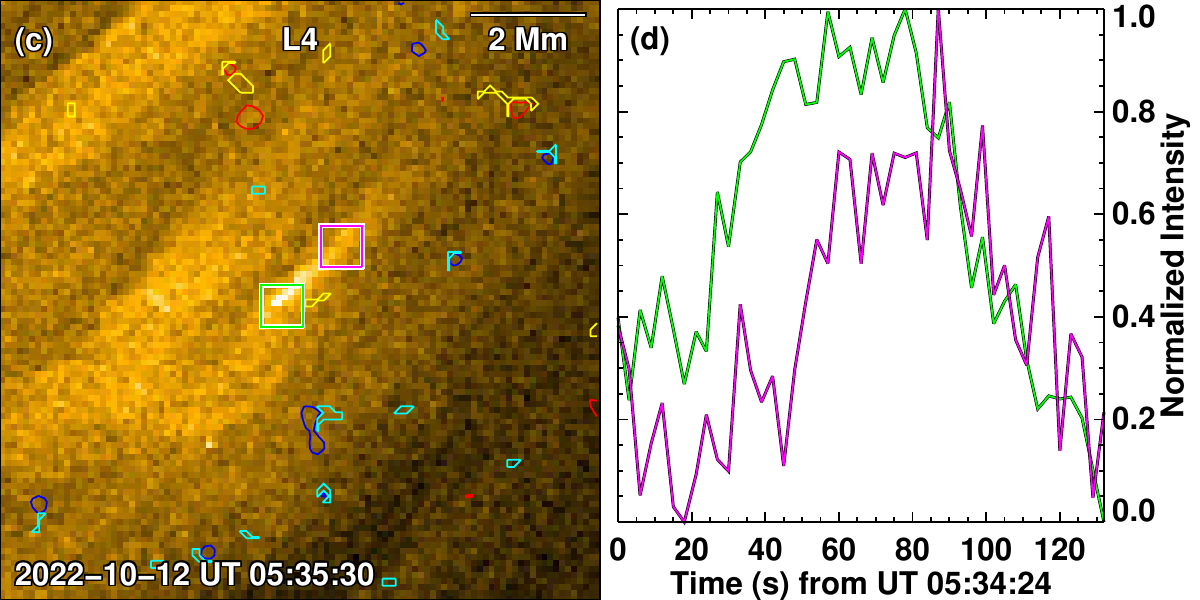}
\includegraphics[width=0.49\textwidth]{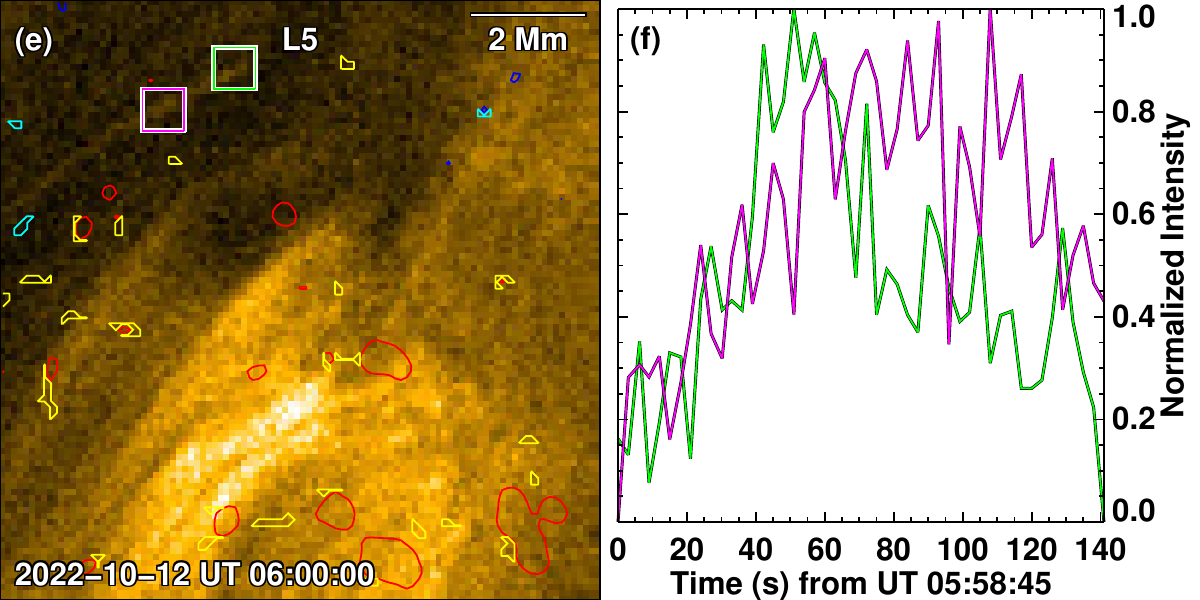}
\includegraphics[width=0.49\textwidth]{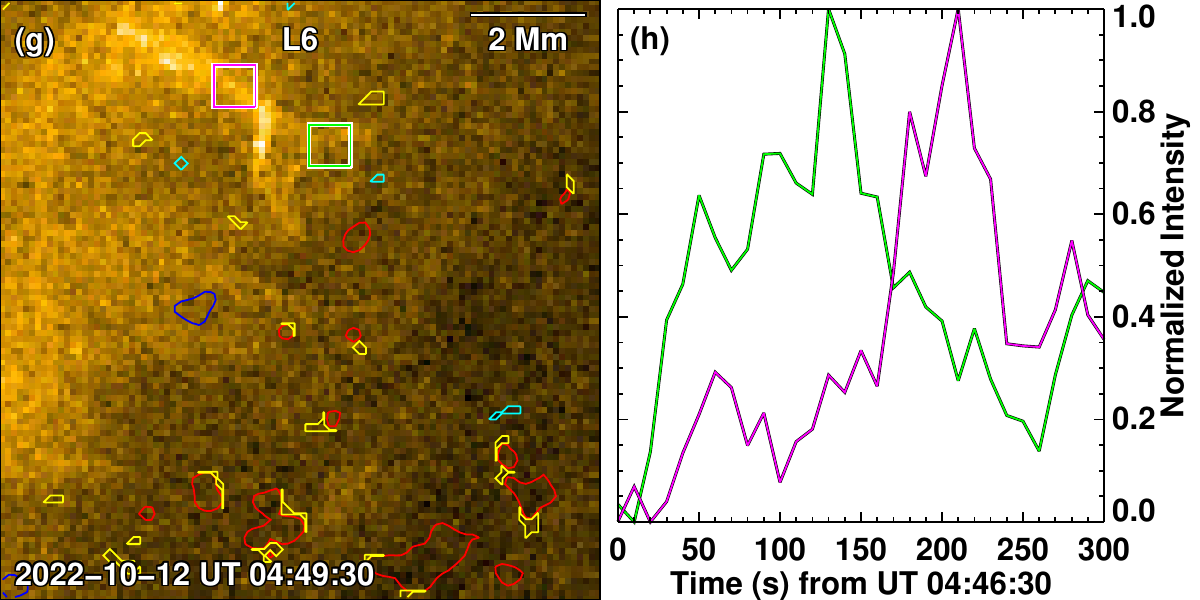}
\caption{Coronal dynamics emerging from weaker-field regions. Panels (a), (c), (e), and (g) display examples of four jet-like events from the weak-field sections of loops, L1, L4, L5, and L6. The green and purple colored boxes are positioned along the parts of these jet-like activity. The magnetic flux density contours have the same meaning as in Fig.\,\ref{fig:over1}. The green and purple light curves in panels (b), (d), (f), and (h) show the average \hri\ intensities from the corresponding boxes overlaid on the left-side panels, as functions of time, normalized to their respective minima and maxima within the plotted time period. These light curves are derived from the images at their native cadence. Sect.\,\ref{sec:obs} and Appendix\,\ref{app:proc} for details.\label{fig:events}}
\end{center}
\end{figure*}

\section{Surface magnetic landscape of the quiet-Sun corona\label{sec:land}}
Many coronal loops or bundles of loops in the considered FOV appear to have one of their footpoints rooted in network magnetic patches. As such, the central part of the \hri\ FOV displayed is dominated by loops connected to the elongated network patch (see also the time-averaged maps in Fig.\,\ref{fig:avg}). By following the animation associated with Fig.\,\ref{fig:over1}, we visually identified some coronal loop systems, labeled L1--L8. For example, in the displayed instance in Fig.\,\ref{fig:over1}, loop features ending in boxes L3, L5, and L7 are evident.

We present a closer look at one of the footpoint regions or segments of these loop bundles and their evolution in Figs.\,\ref{fig:lps1} and \ref{fig:lps2}. Alongside, we also plot the contours of underlying magnetic field patches (as defined in Sect.\,\ref{sec:obs}) near the considered footpoints of these loop segments. For instance, in case of loop L1, only a part of the loop segment is initially seen to be connected to a well-detected positive polarity magnetic patch of size less than 1\,Mm (red contoured region, over 3-$\sigma$ level, closer to the center of Fig.\,\ref{fig:lps1}a). With time, this magnetic patch is no longer distinguishable, while the coronal loop continues to persist (Fig.\,\ref{fig:lps1}b--d). This footpoint region is also bustling with short-living weaker magnetic patches of both positive and negative polarities.

Similar magnetic field and coronal behavior is observed in loops L2--L4. In case of loop L2, the larger positive polarity magnetic field patches show some persistence. But, there is also an increased presence of both well-detected and weaker negative-polarity magnetic field patches, on small spatial scales, at that location (Fig.\,\ref{fig:lps1}e--h). Small-scale magnetic field patches at the bases of loops L3 and L4 are rather scattered and show little persistence over time (Fig.\,\ref{fig:lps1}i--p).

Compared to all the other examples, the loop bundle L5 has one of its footpoints rooted in magnetic field patches over 3-$\sigma$ level, that are persistent. Still, these magnetic patches are flanked by fleeting small-scale mixed-polarity magnetic fields (Fig.\,\ref{fig:lps2}a--d). The segment of loop L6 is again rooted in mixed-polarity footpoint region. It is striking to note that a brighter portion of the footpoint region is devoid of clear underlying magnetic structures (regions outlined by ellipses in Fig.\,\ref{fig:lps2}f--g).

Loop features L7 and L8 are more dynamic. In case of loop bundle L7, it is evident from Fig.\,\ref{fig:over1} that its conjugate footpoint is rooted in the predominantly negative polarity magnetic network patch. Therefore, it is natural to expect that the region bounded by box L7 is composed of positive polarity magnetic concentrations. However, a zoom into this region reveals the presence of well-detected patches of negative polarity magnetic fields directly at the footpoint regions along with well-detected positive polarity fields, in addition to weaker mixed-polarity fields (Fig.\,\ref{fig:lps2}i--l). Time-averaged magnetic field map also reveals the persistence of negative polarity magnetic elements in box L7 over the course of 2.25 hours of observations, along with positive polarity magnetic elements surrounding them (see box L7 in Fig.\,\ref{fig:avg}). Loop L8 is fully contained within box L8 and is related to an eruption or larger-scale reconfiguration of magnetic fields in that location as inferred from the rapidly varying coronal emission (see Fig.\,\ref{fig:over1} and the associated online animation). This loop bundle shows two clear footpoints on either side, but the underlying magnetic field distribution is quite random without a clear bipolar structure (Fig.\,\ref{fig:lps2}m--p).

We find that while one footpoint of a QS coronal loop bundle might end in network magnetic field patches, the opposite end or the conjugate footpoint regions are complex, and are riddled with weaker and small-scale mixed polarity magnetic fields that evolve on timescales of 5\,minutes or less. But the question is whether these small-scale, often weaker magnetic features play any active role in the dynamics of the overlying corona.

\section{Coronal dynamics emerging from weaker field regions\label{sec:jets}}
To demonstrate the active role of these weaker small-scale fields at QS coronal loop footpoints, we present examples of jet-like eruptive activity emerging from these regions in four of the studied loop bundles (Fig.\,\ref{fig:events}). Near the footpoints of loop L1, we observed a slender propagating coronal disturbance that traversed a distance of at least 5\,Mm from its source region (green to purple box locations in Fig.\,\ref{fig:events}a). The propagating nature of the disturbance is evident from the localized intensity fluctuations: the source region (green light curve) exhibits intensity enhancements before the disturbance showing signatures farther away (purple light curve). The overall disturbance, as inferred from the increase and decrease of localized intensity fluctuations lasts for about 50--100\,s. At this footpoint, we found impersistent positive polarity magnetic elements. There are no well-detected negative polarity magnetic patches, per our definition in Sect.\,\ref{sec:obs}, within at least 1\,Mm separation from the jet footpoint, at the time of the activity.

Events closer to the footpoint regions of loops L4 and L5 emerge over locations with seemingly no well-detected underlying magnetic features (Fig.\,\ref{fig:events}). The jet event from L6 region even exhibits a 1--2\,Mm wide dome like structure with an elongated spine feature and appears to emerge from weaker magnetic field regions, if any.

We detect a continually changing magnetic landscape at the feet of coronal loops. The data also reveal that the small-scale (weaker) magnetic fields act not simply as passive photospheric endpoints of the corona, but might play a role in generating disturbances that propagate into the higher layers.

\section{Discussion\label{sec:disc}}
Our SO/PHI data show that almost all the weaker field patches including the more isolated well-detected field patches are captured only once in any given single frame of the 300\,s cadence series. Obviously, lifetimes of these patches must then be below this interval. Moreover, in our case, SO/PHI scanned the Fe\,{\sc i}\,617.3\,nm line in about 63\,s, i.e., each magnetogram is a temporal average over this period. Therefore, the shortest-lived structures will appear in only parts of the raw data and they would not show up in the magnetograms. Indeed, {\textsc{Sunrise}} observations with 33\,s cadence and with nearly two times better spatial resolution than SO/PHI (when close to the perihelion), suggest that the median lifetime of weaker internetwork magnetic features is only 66\,s \citep[][]{2017A&A...598A..47A}. Given that this measurement in itself might have been limited by the cadence of {\textsc{Sunrise}} observations, there could be weaker surface magnetic features on the Sun with still shorter lifetimes.  

Interestingly, our \hri\ observations do reveal clear spatial variations in the EUV emission patterns and apparent motions on timescales as short as 30\,s (see movie associated with Fig.\,\ref{fig:over1}). Except for the case L8 that formed during an eruption, each of the other loop bundles as a whole persist for longer timescales, ranging from a few 10\,minutes to the entire duration of our observations. But the individual strands within these bundles exhibit intensity variations and evolve on shorter timescales of less than 5\,minutes as displayed in Figs.\,\ref{fig:lps1} and \ref{fig:lps2}. While some of these variations could be related to the general thermal and density changes in the EUV emitting plasma, some could be directly in response to the rapidly changing surface magnetic landscape.

Latter of these EUV Variations, that are potentially linked to the rapidly changing magnetic landscape, can be seen as dynamic events\footnote{The high spatio-temporal resolution offered by \hri led to the studies of a variety of small-scale dynamic and transient coronal events \citep[e.g.,][]{2021A&A...656L...4B,2021A&A...656L..13C,2023Sci...381..867C,2021ApJ...918L..20H,2022A&A...660A.143K,2021A&A...656L..16M,2022A&A...664A..28M,2021ApJ...921L..20P,2023ApJ...943...24P,2022ApJ...929..103T}.} emerging from the weaker field regions. The current understanding is that the type of jet activity discussed in Fig.\,\ref{fig:events} are driven by interaction and reconnection of a parasitic polarity magnetic element with a dominant, opposite-polarity magnetic concentration, during surface flux emergence or cancellation events \citep[][]{2021RSPSA.47700217S}. Here we cannot supply evidence for such reconnection events, though. But the lack of obvious (well-detected) small-scale parasitic polarity fields at the location of these jets suggests that they are not being resolved by our magnetic field observations or that they evolve on timescales shorter than 5\,minutes. Given the median lifetime of 1.1\,minutes of the internetwork magnetic features, the latter is rather likely.

The higher-resolution {\textsc{Sunrise}} observations recorded smallest magnetic fluxes of $9\times10^{14}$\,Mx \citep[][]{2017A&A...598A..47A,2017ApJS..229...17S}. In our data for comparison, the smallest value of the magnetic flux for well-detected fields from a single pixel, that SO/PHI can measure is about $3.5\times10^{15}$\,Mx. Still weaker magnetic features will obviously remain undetected in our data. Surface magnetic fields on the Sun follow a power-law distribution over seven orders of magnitude in flux (10$^{16}$--10$^{23}$\,Mx), with an index of $-2.69$ \citep[][see also \citealt{2017A&A...598A..47A} and \citealt{2017ApJS..229...17S}]{2011SoPh..269...13T}. Assuming that the same power-law index continues to lower fluxes of $10^{15}$\,Mx,
magnetic elements with flux contents in the range of 10$^{15}$--10$^{16}$\,Mx will emerge at 4 times higher rate compared to those in the range of 10$^{16}$--10$^{20}$\,Mx \citep[c.f. Eq. 3 in][]{2011SoPh..269...13T}. This implies that magnetic elements with flux content below our detection limit (i.e., $<3.5\times10^{15}$\,Mx) still could be crucial for coronal loop structuring and atmospheric dynamics.

In the selected loop bundles, we have mainly focused on the footpoints rooted in the generally weaker field regions. For instance, the conjugate footpoints, i.e., the opposite ends of loop bundles L1--L7 (Fig.\,\ref{fig:over1}) end in network magnetic field patches. There, too, the presence of dynamic, weaker mixed polarity fields is clear. These features might be related to the internetwork elements that emerge closer to or are swept into the network magnetic concentrations \citep[][]{2014ApJ...797...49G}. Their interaction with the network magnetic features might give rise to enhanced atmospheric activity \citep[][]{2019Sci...366..890S,2022ApJ...924..137W,2023ApJ...944..171B}.

Previously, the existence of coronal magnetic connections to photospheric weak field regions could only be inferred and argued through magnetic field extrapolations of synthetic magnetograms, representative of the QS \citep[][]{2003ApJ...597L.165S,2006A&A...460..901J} or observed magnetic fields \citep[][]{2010ApJ...723L.185W,2013SoPh..283..253W}.
Here we observationally demonstrated that QS coronal loops do connect to surface regions with rapidly varying, weaker magnetic fields. Moreover, we also have shown evidence for coronal dynamics emerging from the underlying weaker field regions.

Here we implicitly assume that the EUV segments of a loop bundle connect nearly vertically to the underlying magnetic fields. In addition, the EUV segment of the loop bundle need not necessarily trace the full length/arc of the magnetic connectivity. While the coronal segment of the loop around 1\,MK emitting in the EUV is detected by \hri, the instrument will not detect/capture the lower parts of the loop filled with denser and cooler chromospheric material. In such a case, it could very well be that the apparent EUV endpoints of a coronal loop do not actually point to the surface magnetic footpoints. Then the magnetic fields threading a loop could bifurcate closer to the surface and connect to other distinct flux patches \citep[][]{2002ApJ...576..533P} or extend further along and connect to surface regions that are laterally separated from the assumed footpoints. Even should this be the case, our key interpretation of the presented observations still stands, namely that the underlying magnetic patches in the vicinity of the assumed footpoints of the loop bundles are still small-scale dynamic features possessing generally weaker fields. 

On a data-processing level, there could be residual offsets between the coronal images and photospheric magnetic field maps. We assess that the alignment accuracy is better than 1--2\,Mm, but cannot be worse than 5\,Mm in any given point of the FoV (i.e. the scale shown in Fig.\,\ref{fig:over1}). Thus there could be some offsets unaccounted for between the coronal ends of the loops and the surface magnetic footpoints. Again, as discussed above, in their immediate vicinity at the surface, even up to $\pm$5\,Mm separation, the magnetic field remains qualitatively similar. Therefore any minor offsets in the alignment will not affect our inferences.

The morphology of the \hri\ loop bundles generally resembles that of the QS coronal bright points around 1--2\,MK \citep[][]{2019LRSP...16....2M}. Although the thermal response of \hri\ has a peak around 1\,MK, it also has a secondary contribution at lower temperatures of $\sim$0.3\,MK from O\,{\sc vi} \citep[][]{2021A&A...656L...7C}, sampling emission from the transition region. This means that the studied loop bundles could actually be the transition region counterparts of coronal bright points \citep[][]{2008ApJ...681L.121T}. Irrespective of their exact thermal characteristics, the evident dynamic nature of the magnetic fields at their footpoints would still impact the evolution of any cooler/hotter atmospheric sections of these loops.

\section{Implications for the quiet-Sun coronal heating\label{sec:conc}}

Our observational findings have important implications for coronal heating models. For instance, wave-based heating models \citep[][]{2020SSRv..216..140V} do not yet consider the Poynting flux and wave energy injection into the upper atmosphere due to the emergence of small-scale magnetic features, their rapid variation, interaction with other small-scale features and possible recycling.

Another traditional model of coronal heating is based on the energy dissipated in the process of untangling of magnetic fields braided by slow stressing of footpoints \citep[][]{1988ApJ...330..474P,2006SoPh..234...41K,2013Natur.493..501C,2022A&A...667A.166C}. In this scenario, a typical QS coronal bright point with loop length of 10\,Mm might reach a steady state between the energy injected into the corona through slow footpoint motions and energy losses through radiation and conduction, on timescales of 1.4\,hours \citep[][]{1988ApJ...330..474P}. Indeed, this timescale is close to the coronal magnetic field recycling timescale, as inferred from the analysis based on low-resolution SOHO/MDI magnetograms \citep[][]{2004ApJ...612L..81C}. But the rapid spatio-temporal variations we observed mean that magnetic fields underlying even the QS loops are even more complex than previously thought, further implying that coronal fields might be subjected to rapid tangling and untangling, on shorter timescales of only a few 100\,s or even less. In essence, the coronal segment of a loop might not retain its surface magnetic identity, for a slow build up of energy in the corona over a 1.4\,hour timescale, to begin with. Then how the processes of rapid untangling and the slow build up of energy compete in generating coronal heating and corresponding observational signatures in the braiding scenario, needs to be evaluated.

At the same time, complex magnetic fields might reconnect in the lower atmosphere and release the energy already in the chromosphere and transition region at the coronal base, leading to hotter plasma filling the coronal loops \citep{2007ApJ...659.1673A}. If these variations are sustained by frequent flux emergence and cancellation events in the photosphere \citep[][]{2017ApJS..229...17S}, they could power the overlying coronal loops through magnetic reconnection \citep[][which is indeed observed in some active region loops, e.g., \citealt{2014ApJ...795L..24T}, \citealt{2017ApJS..229....4C,2020A&A...644A.130C}]{2018ApJ...862L..24P}.

As one can see in Figure~\ref{fig:over1}, the correspondence between QS coronal loop bundles and photospheric magnetic field patches is generally clear on larger scales. However, such a correspondence for each individual small-scale loop is far from certain. Even in cases with an obvious overall correspondence, like regions L3 and L5 (see Figures~\ref{fig:lps1}i--l and \ref{fig:lps2}a--d), it is not evident which small-scale loop connects to which photospheric magnetic field patch. On the one hand, this may indicate that the heating that leads to the EUV emission is located directly in the corona and not in the transition region or even lower. On the other hand, this could also point to the role of unresolved mixed polarity magnetic elements at the base of coronal loops.

Indeed, high-resolution simulations of the surface magnetoconvection show the existence of closely spaced mixed polarity magnetic fields in intergranular lanes \citep[][]{2007A&A...465L..43V,2014ApJ...789..132R}. Depending on their relative strengths and spatial separation, their observable effect, within the resolution element, could be diminished in our SO/PHI observations. A direct indication of such an effect could be seen in our \hri\ data where the loops apparently root in regions with little surface magnetic flux (see Fig.\,\ref{fig:lps2}). Therefore, it is imperative that the role of (weaker) small-scale surface magnetic fields along with their recycling, emergence and cancellation need to be carefully evaluated in the processes of mass and energy transfer into the corona \citep[see also][]{2022SoPh..297..129W}. It is also interesting to note that despite the dynamism of surface magnetic fields, there is a clear persistence in the overlying coronal loops in the course of about 2.25\,hours (Appendix\,\ref{app:avg}; Fig.\,\ref{fig:avg}). This kind of temporal persistence could be an additional constraint on different heating models.

Using unprecedented photospheric and coronal coordinated observations with SO/PHI and EUI instruments onboard Solar Orbiter, we have demonstrated the direct and crucial role of small-scale magnetic fields in the structuring of the QS solar corona. The rapidly varying yet dominant weaker magnetic flux patches could be important for the overlying atmospheric dynamics and eventually for the powering of the solar corona itself. 

%
%

\begin{acknowledgements}
The authors thank the anonymous referee for constructive comments that helped improve the manuscript. L.P.C. gratefully acknowledges funding by the European Union (ERC, ORIGIN, 101039844). Views and opinions expressed are however those of the author(s) only and do not necessarily reflect those of the European Union or the European Research Council. Neither the European Union nor the granting authority can be held responsible for them. Solar Orbiter is a space mission of international collaboration between ESA and NASA, operated by ESA. We are grateful to the ESA SOC and MOC teams for their support. The EUI instrument was built by CSL, IAS, MPS, MSSL/UCL, PMOD/WRC, ROB, LCF/IO with funding from the Belgian Federal Science Policy Office (BELSPO/PRODEX PEA 4000112292 and 4000134088); the Centre National d’Etudes Spatiales (CNES); the UK Space Agency (UKSA); the Bundesministerium für Wirtschaft und Energie (BMWi) through the Deutsches Zentrum für Luft- und Raumfahrt (DLR); and the Swiss Space Office (SSO). The German contribution to SO/PHI is funded by the BMWi through DLR and by MPG central funds. The Spanish contribution is funded by AEI/MCIN/10.13039/501100011033/ and European Union ``NextGenerationEU''/PRTR'' (RTI2018-096886-C5,  PID2021-125325OB-C5,  PCI2022-135009-2, PCI2022-135029-2) and ERDF ``A way of making Europe''; ``Center of Excellence Severo Ochoa'' awards to IAA-CSIC (SEV-2017-0709, CEX2021-001131-S); and a Ramón y Cajal fellowship awarded to DOS. The French contribution is funded by CNES. A.N.Z. thanks the Belgian Federal Science Policy Office (BELSPO) for the provision of financial support in the framework of the PRODEX Programme of the European Space Agency (ESA) under contract number 4000136424.
\end{acknowledgements}

%
%

\appendix
\section{Data processing and alignment\label{app:proc}}

\begin{figure*}
\begin{center}
\includegraphics[width=0.6\textwidth]{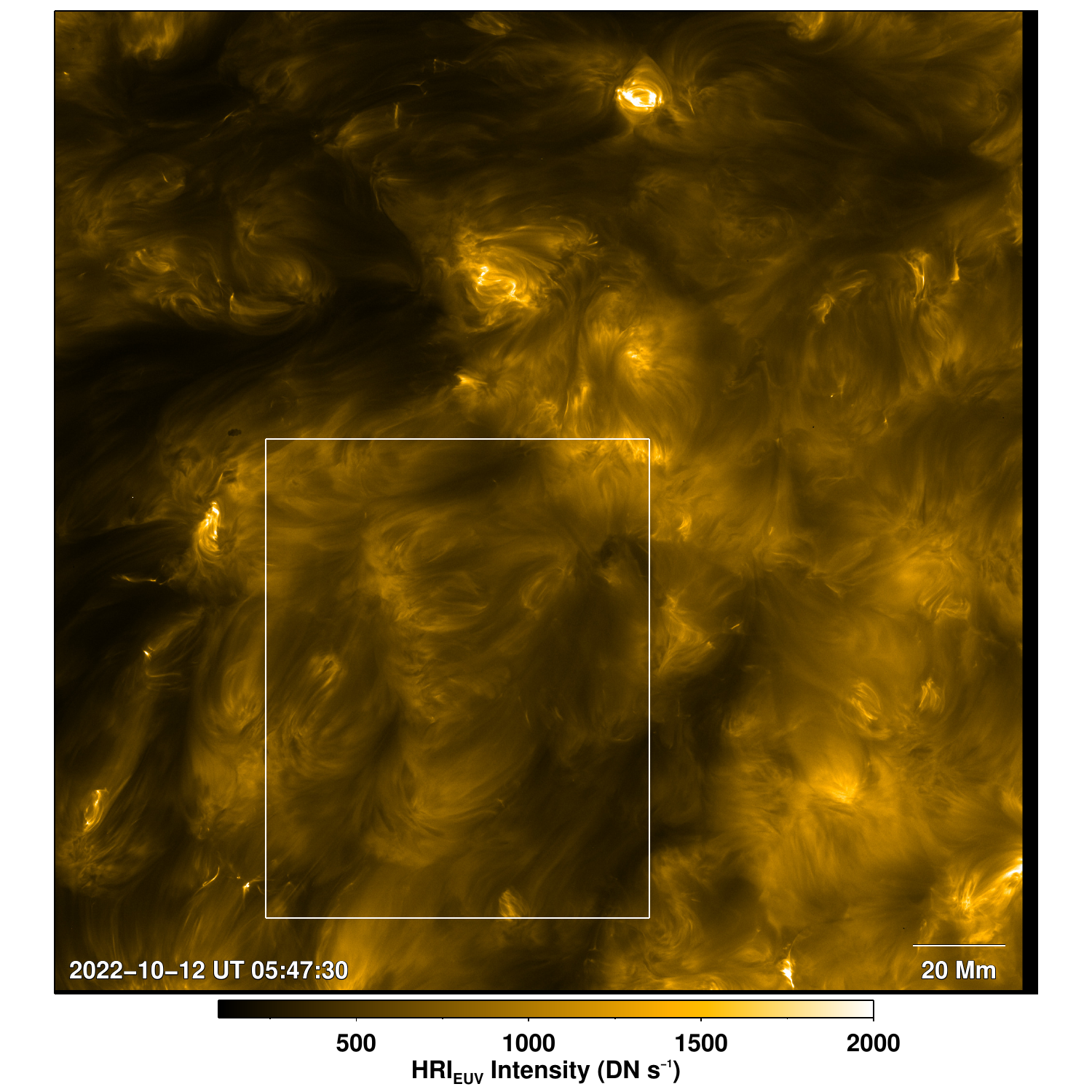}
\caption{Quiet-Sun corona and plasma loops. The snapshot shows the full FOV of \hri. The white box marks a partial FOV that we presented in Fig.\,\ref{fig:over1}. See Sect.\,\ref{sec:obs} and Appendix\,\ref{app:proc} for details.\label{fig:euifov}}
\end{center}
\end{figure*}

The EUI and SO/PHI data were further processed as follows. First we co-aligned all the EUI data to the first image of the time sequence using a cross-correlation technique described in \citet{2022A&A...667A.166C}. We used these aligned level-2 EUI data at their native cadence to compute the light curves plotted in Fig.\,\ref{fig:events}. Next, the aligned EUI data (composed of sub-sequences with 10\,s, 3\,s, and 10\,s cadences) were further processed to a uniform effective cadence of 30\,s, by temporally averaging an appropriate number of images.

We also created an array of time-stamps with 30\,s interval to match the 30\,s effective cadence sequence. Here the start time of the new time-stamps is the same as the time-stamp of the first EUI image in the original L2 sequence.

The 10\,s cadence sub-sequences were recorded with an Aluminum filter placed in front of the detector, while the 3\,s cadence sub-sequence images were recorded with no filter. Consequently, per exposure time, roughly 1/0.6 more photons were being recorded by the detector with no filter compared to the Aluminum filter case. This means that the exposure normalized intensities in the 3\,s cadence images are brighter by a factor 1/0.6 compared to the images from the 10\,s cadence sub-sequences. For display purposes, we account for these intensity differences by further normalizing the 30\,s effective cadence data with the corresponding mean intensity at each time step, computed over the field of view displayed in Fig.\,\ref{fig:over1}. Moreover, we applied an unsharp mask filter to sharpen the images displayed in Figs.\,\ref{fig:over1}, \ref{fig:lps1}, \ref{fig:lps2}, and \ref{fig:events}. The full FOV of \hri\ displayed in Fig.\,\ref{fig:euifov} is also from the 30\,s effective cadence data, but with no unsharp mask filter applied.

The SO/PHI magnetic field vector is derived using the Milne-Eddington approximation under the assumption that the Fe\,{\sc i}\,617.3\,nm spectral line is formed in local thermodynamic equilibrium \citep[LTE;][]{2020A&A...642A..11S}. However, this line has been shown to be affected by non-LTE conditions in the solar atmosphere. Ignoring these effects might result in an underestimation of magnetic fields \citep[][]{2023A&A...669A.144S}. Such non-LTE effects, however, are not considered in this study.

We re-scaled the SO/PHI maps to the slightly smaller pixel scale of \hri, that resulted in 1041$\times$1041\,pixels data. These re-scaled SO/PHI maps were all co-aligned to the map recorded closest to the middle of the time sequence (around UT\,05:47). Then the first image from the 30\,s effective cadence \hri\ time sequence was visually aligned to the closest-in-time re-scaled SO/PHI map. Our visual alignment was primarily guided by a prominent elongated magnetic network feature consisting of negative polarity magnetic patches (enclosed by a green ellipse in Fig.\,\ref{fig:over1}a). Conspicuous coronal structures associated with this magnetic network are seen in the \hri\ data (green ellipse Fig.\,\ref{fig:over1}b). Keeping re-scaled aligned SO/PHI data as reference, this visual alignment was achieved by extracting the \hri\ sub-FOV with $(x,y)=(340,140)$ as origin (Fig.\,\ref{fig:euifov}). This alignment also naturally led to the alignment of other clearer loop systems to their underlying magnetic fields (e.g., Loop bundle associated with L5 and a bipolar loop structure at the bottom of the frame, adjacent to L8). Overall, we assess that the alignment accuracy is better than 1--2\,Mm. Figure\,\ref{fig:over1} then represents a still smaller sub-FOV covering $x=$100--899 pixels and $y=$17--1016 pixels in the re-scaled SO/PHI data and the corresponding extracted \hri\ sub-FOV. Finally, we padded the SO/PHI time sequence by populating the temporal dimension with the corresponding maps that are nearest in time to the 30\,s effective cadence EUI sequence.

\section{Average Maps\label{app:avg}}
\begin{figure*}
\begin{center}
\includegraphics[width=\textwidth]{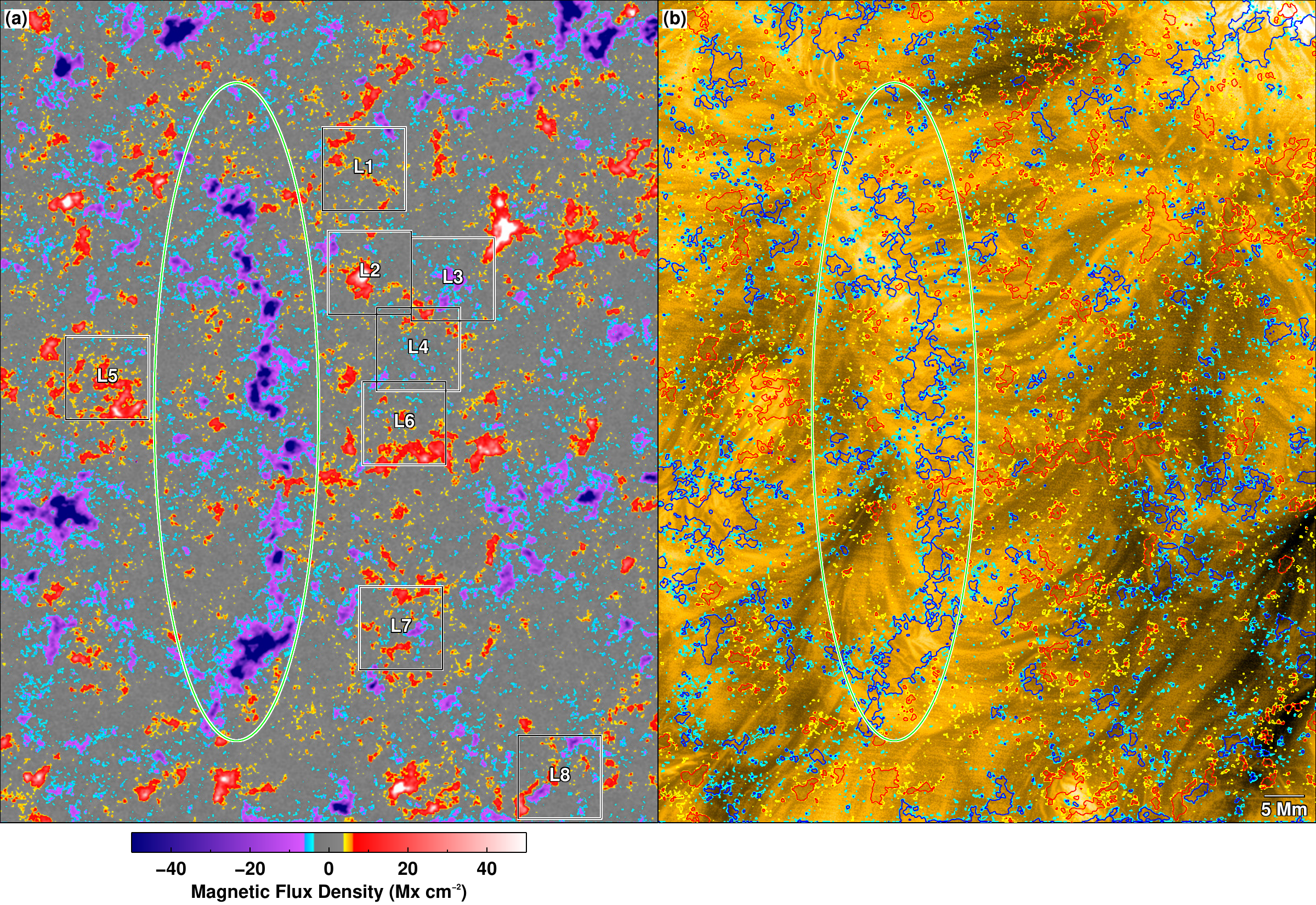}
\caption{Persistent surface magnetic and coronal loop features. Time-averaged SO/PHI magnetic flux density (panel a) and the \hri\ coronal (panel b) maps are displayed. Green ellipse and boxes L1--L8 have the same meaning as in Fig.\,\ref{fig:over1}. The red (positive) and blue (negative) contours outline magnetic concentrations with flux densities above 6\,Mx\,cm$^{-2}$ (i.e., 3-$\sigma_\textrm{avg}$), while the yellow (positive) and cyan (negative) contours mark regions with flux densities between 2-$\sigma_\textrm{avg}$ and 3-$\sigma_\textrm{avg}$ level, based on the time-averaged SO/PHI magnetic flux density map. See Appendices\,\ref{app:proc} and \ref{app:avg} for details. \label{fig:avg}}
\end{center}
\end{figure*}

To check the persistence of surface magnetic fields and the overlying QS coronal loops, we created temporal averages of SO/PHI magnetic flux density and \hri\ coronal maps from the respective time sequences spanning the length of the EUI observations (i.e. about 135\,minutes). There are in total 27 SO/PHI magnetic field maps covering this period, which we temporally average. The 1-$\sigma$ noise level in the magnetic flux density in a single map is about 10\,Mx\,cm$^{-2}$. Assuming Poisson statistics, the 1-$\sigma$ noise level in the time-averaged magnetic flux density will reduce by a factor of $\sqrt{27}$, to $\sim$1.9\,Mx\,cm$^{-2}$ (referred to as $\sigma_\textrm{avg}$ for brevity in the following). The averaged SO/PHI map is displayed in Figure\,\ref{fig:avg}a. The prominent elongated magnetic network is evident in this map (encircled by an ellipse; see also Fig.\,\ref{fig:over1}).

There are 1428 \hri\ images in the aligned level-2 EUI native-cadence sequence, which are temporally averaged. We contrast-enhance the time-averaged coronal map by a multi-scale Gaussian normalization technique \citep[][]{2014SoPh..289.2945M} to better visualize the features (Fig.\,\ref{fig:avg}b). Persistent QS coronal loops rooted in the elongated magnetic network are evident in this time-averaged map.

\bibliographystyle{aasjournal}

\end{document}